# THE B850 / B875 PHOTOSYNTHETIC COMPLEX GROUND AND EXCITED STATES ARE BOTH COHERENT

R. H. Squire [1, δ], N. H. March [2], A. Rubio [3], J. R. Ingles [1, †], W. A. Cunningham [1, †]

[δ] Department of Natural Sciences
[†] Undergraduate STEM Student
Department of Chemical Engineering
[1] West Virginia University - Institute of Technology
Montgomery, WV
[2] Department of Physics
University of Antwerp
Groenborgerlaan 171, B-2020
Antwerp, Belgium
Oxford University
Oxford, England
[3] NanoBio Spectroscopy Group and ETSF Scientific Development Centre and DIPC
Department of Materials Science, Faculty of Chemistry
University of the Basque Country  UPV/EHU
Centro Joxe Mari Korta, Avenida de Tolosa, 72,
E-20018 Donostia-San Sebastian, Spain

**Abstract**

A bacterial photosynthetic light-harvesting complex (PLHC) absorbs a photon and transfers this energy almost perfectly *at room temperature (RT)* to a Reaction Center (RC), where charge separation occurs.  While there are a number of possible light absorbers involved in this process, our focus is the B850 and B875 complexes.  We propose that the dominant feature of the ground states in the B850 ring and the B875 "open chain" are pseudo one-dimensional metals due to each bacteriochlorophyll a (BChl) containing a coordinated magnesium ion – Mg(2+).  The Mg ion structure undergoes a static Peierls' distortion that results in "symmetry breaking" that changes the even spacing of the Mg/BChl molecules comprising the chains to the experimentally observed Mg/BChl "dimers".  The results are charge density waves (CDW, one for each type of the two complexes) that result in an energy gap in the single-particle electronic spectrum and coherent phonons spanning the entire rings.  The *ground state* CDW's seem to have two functions: the first is to form a stable optical platform and the second is to suppress radical formation and energy dissipation of the coherent excited state by creating single-particle energy gaps.  After *excitation* by a photon, the B850 exciton delocalizes on the ring; a second photon can form a two-level exciton-polariton that could be an alternative explanation for the splitting of the B850 "exciton band".  The coherent polariton formed could actively participate in "uphill" electronic energy transfer (EET) [1]. Additionally, we suggest other possible energy storage mechanism and entanglement possibilities.  We suggest experimental studies to clarify these proposals.

**1a. Introduction.** In anaerobic photosynthetic prokaryotes (purple bacteria), the process of light capture and transfer is highly efficient. Our emphasis is on capture of solar energy and its nearly perfect transfer to the reaction center (RC or P870). In particular, a purple bacteria called Rhodopseudomonas acidophila has a photosynthetic unit comprised of two pigment-protein light-harvesting (LH) complexes called LH1/RC and LH2. These LH's have similar molecular structures comprised of bacteriochlorophyll components (BChl a, designated BChl hereafter) that are comprised of circular conjugated arrays with a coordinated magnesium ion, Mg(2+), Fig 1a. The RC-LH1 is surrounded by eight LH2's and the atomic structures and orientations and

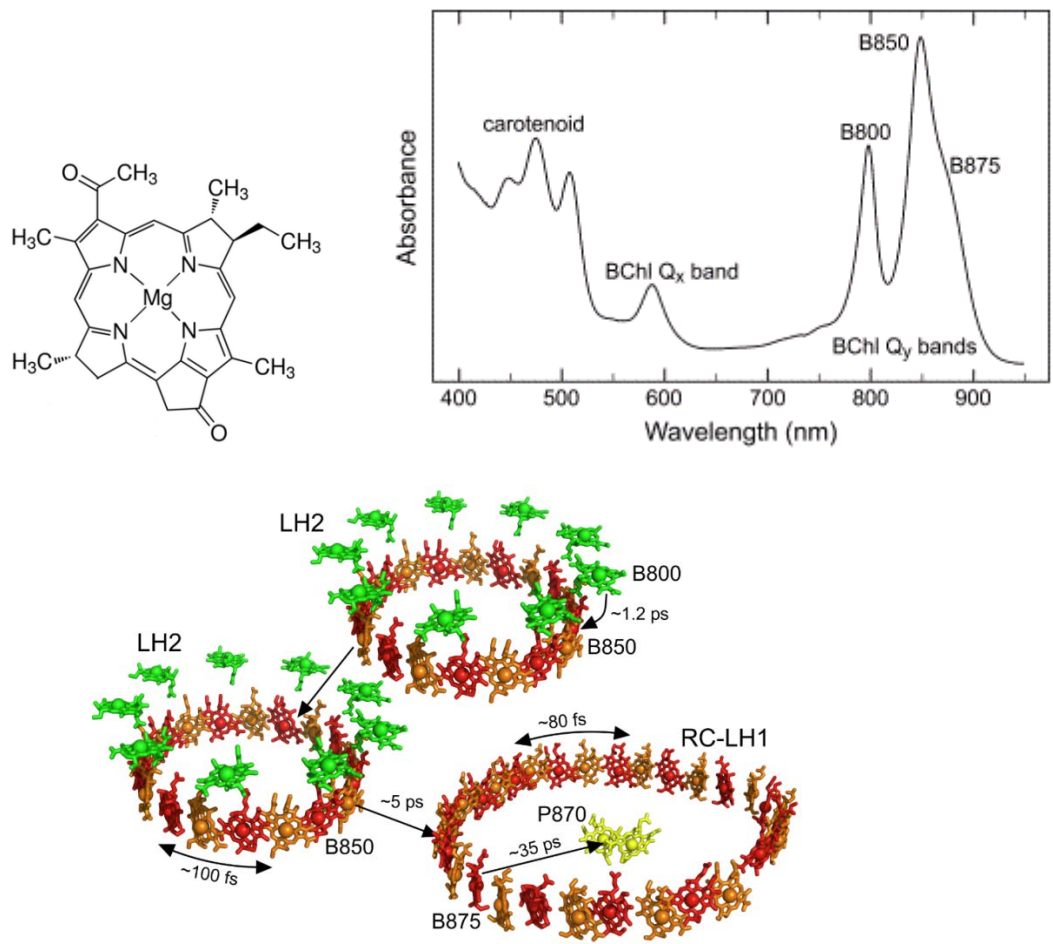

Figure 1a (top left) A prototypical bacteriochlorophyll a (BChl) (Wehlitz); (right) Spectrum of BChl complexes B800 (green below), B850 and B875 (red and yellow). The Soret bands are located between 300 – 400 nm. 1b. Graphical illustration of crystal structure coordinates for light harvesting (LH) complexes 1, 2 and the reaction center (P870). The orange, red and green structures are comprised of the BChl ring structure surrounding a Mg(+2) that is responsible for the charge density wave that generates the "dimer" structure of the B850 and B875 rings (see text). The absorption maxima indicate the complexes; EET rates are shown. Figure courtesy of Professor M. Jones (Bristol).

orientations are known [2], and the associated pigments groups are separated clearly enough so that EET can be measured. The ultimate objective is to transfer energy efficiently to the reaction center (RC) where "charge splitting" takes place to produce a 1.1eV potential that generates energy for the growth and survival of the bacteria.
___________________________________________________________________

**1b. Structure of a Bio-complex.** The RC is surrounded by a peripheral antenna light harvesting / energy transporting complex, LH1, designated B875, which in turn is surrounded by eight B850 rings, each comprising 16 BChl in the ring (orange and red) and 8 BChl molecules above the ring (green). The RC contains 4 modified BChl rings and B875 has 30 BChl rings, so this highly organized, symmetrical complex (sans RC) contains 326 oriented BChl molecules. Each BChl's in the B800, and B850 and B875 rings contains a coordinated Mg(+2) which is the driving force in the latter two structures to create a coherent ground state called a charge density wave (CDW). Initially we will assume that the B800 and other peripheral light absorbing compounds are incoherently connected.

There are two high-resolution crystal structures of the light-harvesting complexes [2, 3], the first examining ten native crystals with an average R factor of 4% (merged) and a completeness of 99.0%. Agreement is excellent, so the usual "soft matter" description does not apply here since the ground state is a rigid CDW (Section 2). The B800 "monomers" are separated from each other by $21.1\overset{o}{A}$ with their molecular plane almost perpendicular to the to the symmetry axis of the LH2 complex; their distance to the nearest B850 is $17.6\overset{o}{A}$. The B850 BChl's are in a "dimer" arrangement with the distance between Mg/BChl centers of $8.9\overset{o}{A}$ with the center of one dimer Mg/BChl to the next dimer center being $9.6\overset{o}{A}$ and all of the molecular planes are parallel to the symmetry axis. The Mg/BChl molecules in the ring are cantered (similar to a turbine blade) relative to their neighbors to create a partial overlapping symmetric ring.

Our interest was piqued by the astounding symmetry of the structures in the rigid crystal as well as the recent report that energy transfer processes between LH2 and LH1 have been experimentally measured in either direction [1]. The ability of both forward and backward energy transfer suggests a novel electronic (EET) process(es). Further, in these bacteria species the amount of light present during the growth of a sample changes the amounts and types of LH2 complexes and peripheral structures synthesized that modifies the spectral intensity, i.e. the bacteria adapt to their environment. Lüer et al have devised an experimental scheme to compare high and low light conditions (HL and LL) that we will discuss [1]. Our aims will be to offer explanations for:

1) Specifically how is energy transferred around LH2 ring in the ground and excited state?
2) Is there a similar transfer mechanism in LH1 and/or RC materials
3) Is the excited state coherent?
4) Is the ground state coherent?
5) What is the mechanism for uphill energy transfer?
6) What are the types of coherence that makes the energy transfer of these steps so efficient?

Section 2 containing a discussion of the mean field CDW, followed by a general discussion of coherence, polaritons and the impact on the PLHC in Section 3. Section 4 discusses the mechanism that supports Josephson Junctions connecting portions of the PLHC along with possible models of energy storage. Section 5 is a brief discussion of possible entanglement and section 6 suggests pertinent experiments, followed by a summary and conclusions (Section 7). **Two appendices** cover the additional details of CDW's and coherence.

**2. Charge Density Waves. a. Introduction.** In the 1930's Peierls discovered that a perfect one-dimensional metal ring would always be unstable to symmetry breaking initiated by electron-electron or electron-phonon interactions that would distort the even spacing of the metal ions into dimers, thereby opening a gap in the energy band that lowered the electronic energy (Fig 2) by $\Delta$ and raised the unoccupied states by $\Delta$ also creating a $2\Delta$ gap. This "Peierls' Distortion" (PD) was the precursor to charge density waves (CDW) [4] and it also occurs in 2 and 3 dimensions. In 1D the Fermi surface consists of just two points, one at $+\vec{k}_F$, the other at $-\vec{k}_F$ that results in an interaction strengthening called "nesting". The sheets of the Fermi surface in 1D allows a

---

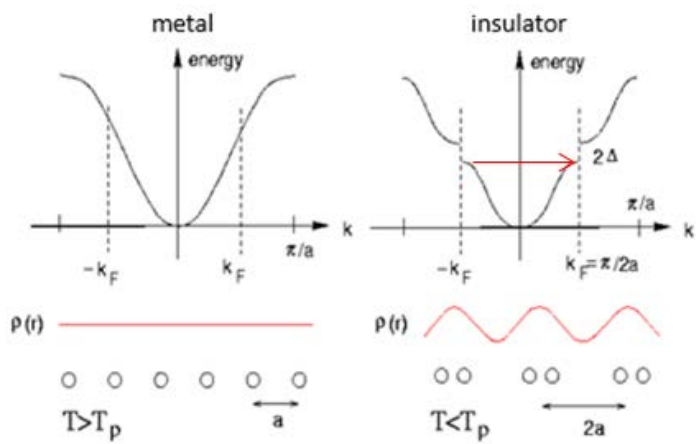

Figure 2. The metal / Mg(2+) band structure (above, left) has a parabolic structure above $T_p$, the Peierls' Distortion temperature. Below $T_p$, (right) the metal atoms group together in a dimer structure that indicates the formation of a CDW and the opening of an energy gap that is comparatively large in our case since the density wave is stable at room temperature. The continuous, coherent (red) phonon wave $\left(q = 2\vec{k}_F\right)$ is undistorted around the B850 complex. The temperature dependence of the mean field CDW equations and the Bardeen, Cooper, Schrieffer (BCS) superconductivity theory [4] are the same, despite some physical differences.

---

density wave to strengthen by a process called "nesting". The B850 ring complex is expected to have perfect nesting (red arrow, Fig 2), as there are no impurities or imperfections present that might localize the coherent phonon mode.

**b. Mean Field Description of PD/CDW [5, 6].** A PD/CDW can be modeled by a 1D free electron gas coupled to an underlying chain of metal ions. A Hamiltonian for this process should contain terms for the electron gas, ionic lattice vibrations, and an electron-phonon coupling.

**Lattice vibrations** are described by the Hamiltonian below,

$$H_{ph} = \sum_q \left\{ \frac{P_q P_{-q}}{2M} + \frac{M\omega_q^2}{2} Q_q Q_{-q} \right\} \tag{1}$$

where M is the mass, $\omega_q$ are the normal coordinates, and $Q_q$, $P_q$ are the standard normal coordinates and conjugate momenta of the atom motions, respectively. Rewriting the Hamiltonian in second quantization notation results in

$$H_{ph} = \sum_q \hbar\omega_q \left( b_q^\dagger b_q + \frac{1}{2} \right) \tag{2}$$

where $b_q$ and $b_q^\dagger$ are the annihilation and creation operators for phonons with wavevector q.

$$Q_q = \left( \frac{\hbar}{2M\omega_q} \right)^{1/2} \left( b_q + b_{-q}^\dagger \right)$$

$$P_q = \left( \frac{\hbar M\omega_q}{2} \right)^{1/2} \left( b_q^\dagger - b_{-q} \right)$$

In this notation the lattice displacement is (N is the number of lattice sites per unit length):

$$u(x) = \sum_q \left( \frac{\hbar}{2NM\omega_q} \right)^{1/2} \left( b_q + b_{-q}^\dagger \right) e^{iqx} \tag{3}$$

The **electron-vibration interaction** we will use assumes that the potential V depends only on the distance from the equilibrium lattice position resulting in

$$H_{el-vib} = \sum_{k,k',l} \langle k | V(r-l-u) | k' \rangle a_{k'}^\dagger a_k$$

$$= \sum_{k,k',l} e^{i(k-k')(l+u)} V_{k-k'} a_{k'}^\dagger a_k$$

Here l is the equilibrium atom position, u is the distance from equilibrium, and $V_{k-k'}$ is the Fourier transform of a single atom potential. Approximating for small displacements

$$e^{i(k'-k)u} \approx 1 + i(k'-k)u = 1 + iN^{-1/2}(k'-k)\sum_q e^{iql} u_q$$

Ignoring the interaction of the electrons with ions in their equilibrium positions and focusing on effects at $k_F$ distant from the Brillouin boundaries, we have

$$H_{el-vib} = iN^{-1/2} \sum_{k,k',l,q} e^{i(k'-k+q)l} (k'-k) u_q V_{k-k'} a_{k'}^\dagger a_k$$

$$= iN^{-1/2} \sum_{k,k'} (k'-k) u_q V_{k-k'} a_{k'}^\dagger a_k$$

Expressing the interaction completely in second quantization terms,

$$H_{el-vib} = i \sum_{k,k'} \left( \frac{\hbar}{2M\omega_{k-k'}} \right)^{1/2} (k'-k) V_{k-k'} \left( b^\dagger_{k'-k} + b_{k-k'} \right) a^\dagger_k a_{k'}$$
$$= \sum_{k,q} g_q \left( b^\dagger_{-q} + b_q \right) a^\dagger_{k+q} a_k \qquad (4)$$

where g, the coupling constant, is

$$g_q = i \left( \frac{\hbar}{2M\omega_q} \right)^{1/2} |q| V_q \qquad (5)$$

The **electron gas** part in second quantized form is

$$H = \sum_k \varepsilon_k a^\dagger_k a_k \qquad (6)$$

with energy $\varepsilon_k = \hbar^2 k^2 / 2m$ and $a^\dagger_k, a_k$ being the creation and annihilation operators, respectively. Spin degrees of freedom are omitted, so the density of states $n(\varepsilon_F)$ is for one spin direction is

$$n(\varepsilon) = \frac{L}{\pi \hbar} \left( \frac{m_e}{2\varepsilon} \right)^{1/2} = \frac{L}{\pi \hbar v} \quad \text{and velocity,} \quad v = \frac{\hbar k}{m_e}. \qquad (7)$$

A charge density wave in the adiabatic limit assumes the electrons follow the ions instantaneously and it is represented by using the "Fröhlich electron-phonon Hamiltonian" [8],

$$H = \sum_k \varepsilon_k a^\dagger_k a + \sum_q \hbar \omega b^\dagger_q b_q + \sum_{k,q} g_q a^\dagger_{k+q} a_k \left( b^\dagger_{-q} + b_q \right) \qquad (8)$$

**c. Application to One-Dimensional System.** For a 1D electron gas near $\pm k_F$, the dispersion

___

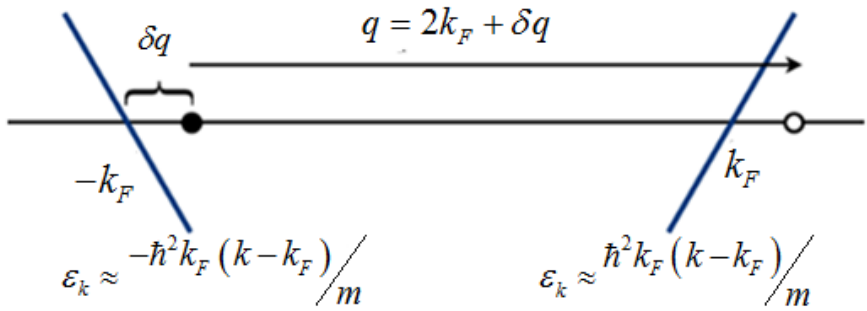

Figure 3. A free fermion gas near the two points that comprise the 1D Fermi surface can be approximated by a linearized dispersion relationship. The particle (filled) and hole (empty) states on opposite sides of the Fermi surface can be coupled by a single $q = 2k_F$ vector.

___

relation is approximated as linear (Fig 3),

$$\varepsilon_k = \hbar v_F (k - k_F) = \pm v_F \delta k \qquad (9)$$

which indicates that *both* the lattice and the electron gas in the (coupled) electron-phonon are unstable as shown below. The effect on the normal vibrational coordinates Q for small amplitude displacement is expressed as

$$\hbar^2 \ddot{Q}_q = -\left[\left[Q_q, H\right], H\right] \tag{10}$$

Since the commutator $\left[Q_q, P_{q'}\right] = i\hbar \delta_{q,q'}$, we have

$$\ddot{Q}_q = -\omega_q^2 Q_q - g\left(\frac{2\omega_q}{M\hbar}\right)^{1/2} \rho_q \tag{11}$$

with $\rho_q = \sum_k a_{k+q}^\dagger a_k$ being the $q^{th}$ component of the electron density (g is assumed independent of q). The second term on the RHS is an effective force constant due to combined electron-vibration interaction that results in a time-independent rearrangement of charge density. So, the response of the electron gas is given by eq (12), and using linear response theory, the

$$\phi(\vec{r}) = \int_q \phi(\vec{q}) e^{i\vec{q}\cdot\vec{r}'} d\vec{q} \tag{12}$$

charge density rearrangement is expressed as an induced charge $\rho^{ind}(\vec{q}) = \chi(\vec{q})\phi(\vec{q})$ where $\chi(\vec{q})$, the Lindhard response function (1D) is ($f_k = f(\varepsilon_k)$ is the Fermi function)

$$\chi(\vec{q}) = \int \frac{d\vec{k}}{2\pi} \frac{f_k - f_{k+q}}{\varepsilon_k - \varepsilon_{k+q}} \tag{13}$$

Applying this expression to eq (9) above and near $2k_F$, the response function becomes

$$\chi(q) = \frac{-e^2}{\pi \hbar v_F} \ln\left|\frac{q+2k_F}{q-2k_F}\right| = -e^2 n(\varepsilon_F) \ln \ln\left|\frac{q+2k_F}{q-2k_F}\right| \tag{14}$$

that diverges indicating that the electron gas is unstable to the formation of a periodically varying electron charge, i.e. when $q = 2k_F$, $\chi(q)$ diverges in 1D (Fig 4).

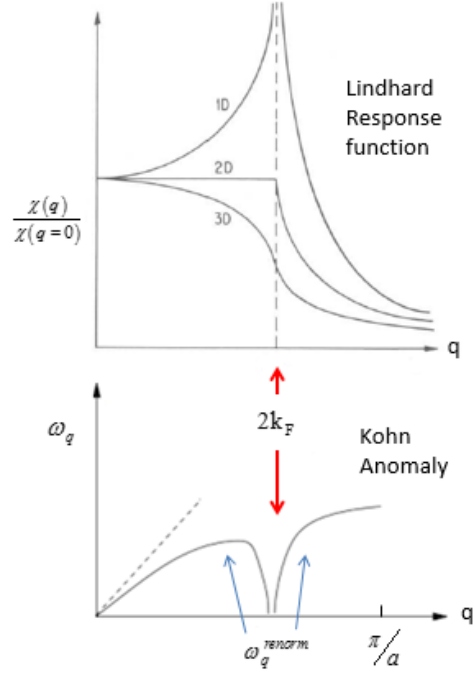

Figure 4. The phonon susceptibility diverges as it approaches $2k_F$, and the renormalized phonon goes to zero and "locks in" the CDW. This renormalization is very temperature dependent; when $\omega_{2k_F}^{renorm} \to 0$, it signals a phase transition to a periodic static lattice distortion of wavelength $\lambda = \pi/k_F$ with a periodic charge modulation.

---

The electron-phonon interaction has generated an effective, but variable force (second term on the rhs of eq. 11) associated with the lattice, but also connecting to the density,

$$\rho_q = \chi(q,T) g \left(\frac{2M\omega_q}{\hbar}\right)^{1/2} Q_q \qquad (15)$$

Using linear response theory leads to the following equation of motion generates a renormalized

$$\ddot{Q}_q = -\left[\omega_q^2 + \frac{2g^2 \omega_q}{M\hbar}\chi(q,T)\right] Q_q \qquad (16)$$

vibrational frequency,

$$\omega_{ren,q}^2 = \omega_q^2 + \frac{2g^2 \omega_g}{M\hbar} \qquad (17)$$

In a 1D model $\chi(q,T)$ has its maximum value at $q = 2\vec{k}_F$, the so-called Kohn anomaly [6] where the reduction (softening) of the vibrational frequency is most significant,

$$\omega_{ren,2k_F}^2 = \omega_{2k_F}^2 - \frac{2g^2 n(\varepsilon_F)\omega_{2k_F}}{\hbar}\ln\left(\frac{1.14\varepsilon_0}{k_B T}\right) \qquad (18)$$

As the temperature is reduced, the renormalized vibration frequency goes to zero which defines a transition temperature where a frozen-in distortion occurs and the phonon mode is "macroscopically" occupied (completely around B850 complex ring) with a non-vanishing expectation value $\langle b_{2k_F} \rangle = \langle b^{\dagger}_{-2k_F} \rangle$ with a complex order parameter defined as

$$|\Delta|e^{i\varphi} = g\left(\langle b_{2k_F} \rangle + \langle b^{\dagger}_{-2k_F} \rangle\right) \quad (19)$$

From the eq. (18) a mean field transition temperature for a charge density wave, $T^{MF}_{CDW}$ (or $T_P$), can be calculated, eq. (20a), and the temperature dependence of the resulting gap equation is *identical to the BCS equation* ($\lambda$ is the electron-phonon coupling constant).

$$k_B T^{MF}_{CDW} = 1.14 \varepsilon_0 e^{-1/\lambda} \quad (20a) \qquad \lambda = \frac{g^2 n(\varepsilon_F)}{\hbar \omega_{2kF}} \quad (20b)$$

The transition temperature for most 1D CDW's known is usually a fraction of the $T^{MF}_{CDW}$ since lattice fluctuations are neglected. Using the BCS temperature relationship for $\Delta$, as can be seen from fig 5, the energy gap (order parameter), $\Delta$, gets larger as the temperature is lowered since [5]

$$2\Delta(T=0) = 3.5 k_B T^{MF}_P$$

Comparing transition temperatures of CDW's with superconductors (SC) in the mean-field approximation, replacing the (SC) Debye energy $\hbar \omega_D$ by the metal Fermi energy $E_F$ results in $E_F / \hbar \omega_D \sim 10-100$, so a Peierls transition temperature above RT is not only possible, the rare-earth tritellurides $(RTe_3)$ are examples [10].

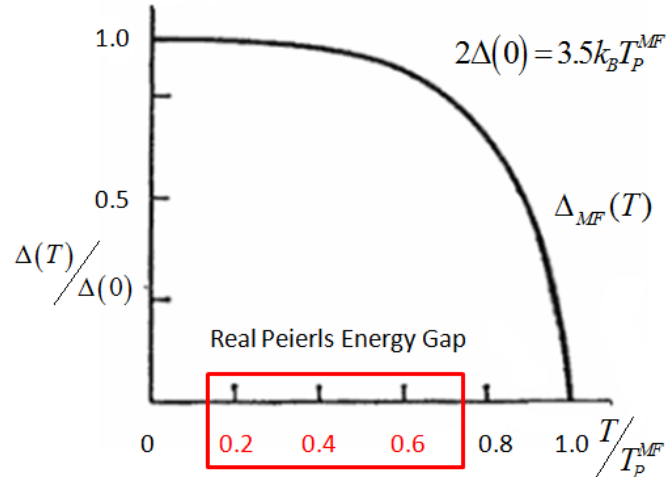

Figure 5. The insulator-metal transition temperatures, $T^{MF}_{CDW}$, for the B850 and B875 complexes are above room temperature; thus, these structures are static since the CDW is "frozen".

The CDW will dramatically affect the density of states (DOS) by creating three gaps. We have discussed the lowest energy single particle energy gap, and there is a band gap generated by reflection symmetry. Lastly, particle-hole symmetry reflects the energy (order) gap so figure xx illustrates the energy diagram with gaps leading to an overall reduction of DOS for, here, the B850 and B875 complexes.

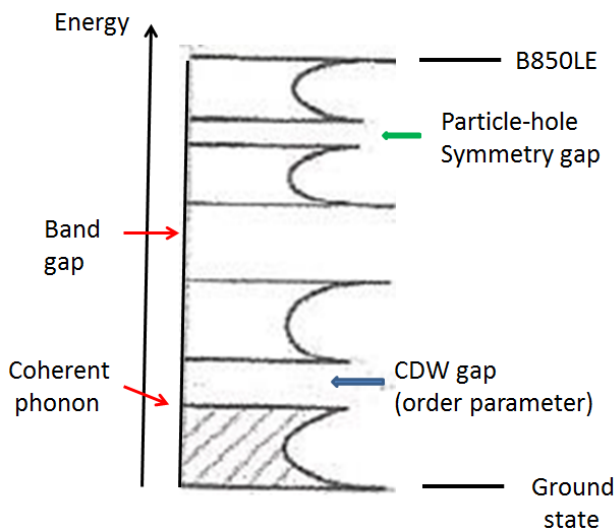

Figure 6. Energy gaps in the CDW spectrum leading to reduction of density of states.

Fröhlich discovered how the propagating combined lattice and electronic charge distorted phase can carry a current as a traveling wave [8]. The wave will travel without attenuation since there is a gap of $2\Delta$ in the energy preventing dissipation. Thus, the conductivity could be infinite because of the CDW's translational invariance. This "broken symmetry" collective mode is essential to the high conductivity. Kuper discusses the thermodynamics associated with a CDW [12]. Detailed discussions of the theories of superconductivity and the excitonic insulator are available [5, 9]. To emphasize, the B850 complex has no impurities or other imperfections, hence, the nesting is perfect. Summarizing thus far, the "dimer" structure in the crystal is an experimental verification of coherence in the ground state B850 ring and B875 structure (mostly a ring) due to the presence of CDW's.

**d. CDW's Ground State Tunneling.** Recently Miller et al [13] have expanded earlier work, particularly by Maki [14] and Bardeen [15] on "depining" CDW trapped by impurities; Bardeen showed that CDW's can tunnel through a potential barrier, similar to Josephson's discovery of cooperative quantum tunneling in superconductors [16]. Coleman broadened the description describe macroscopic quantum tunneling and decay of the "false vacuum" [17] that characterizes instability in a scalar field relative to a lower energy state. We use the theory to describe the propensity of a CDW in a metastable well (B850 excited state) to tunnel to adjacent B850 rings and also to a lower potential well (B875). The event nucleates a bubble of a "true vacuum" contained within a soliton description. The quantum charged $(\pm 2e)$ solitons, delocalized in the transverse directions, could use Josephson tunneling to move to another CDW chain. Following Miller, nucleated droplets of kinks and anti-kinks with a $\pm 2e$ charge can be considered as quantum fluids with quantum delocalization between CDW chains. The model proposed relates a vacuum angle $\theta$ to displacement charge Q between inter-chain contacts $\theta = 2\pi(Q/Q_0)$, where the potential energy of the ith chain is,

$$u_i(\varphi_i) = 2u_0\left[1 - \cos\varphi_i(x)\right] + u_E\left[\theta - \varphi_i(x)\right]^2$$

We ignore the first term, the periodic DW pinning energy, to focus on the quadratic electrostatic energy from the net charge displacement; graphs of u vs. $\theta$ for $\varphi \square 2\pi n$ when the energy is minimized Fig 7. Tunneling is coherent into the next well (chain) via tunneling matrix element T as each parabolic branch crosses the next instability point, $\theta = 2\pi(n+1/2)$. A charge density wave can potentially suppress ionic charges and magnetic field disruption by the Meissner effect ([18], see Section 6 for potential experiment).

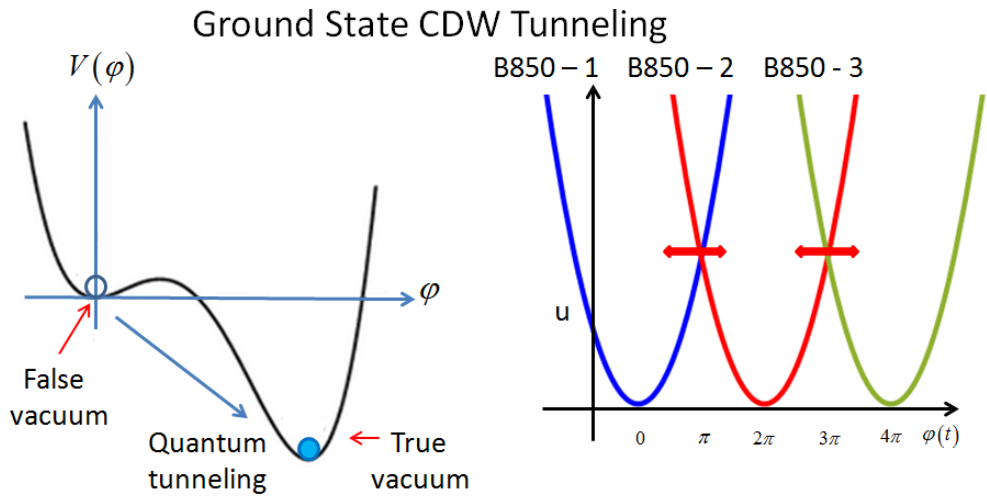

Figure 7. (left) Both tunneling to a lower energy potential well and (right) to a series of identical B850 complex rings can take place. Eight B850 complexes surround the B875-RC complex. If the population changes, say, in B850-2, CDW tunneling will rapidly re-equilibrate the population.

**Why a CDW ground state?** There are several strong possibilities:
1. *Isolate the excited state complexes* with a reduced dissipation by restricting the size of the random heat bath via the two additional CDW energy gaps.
2. Insure rigidity/stability of the B800, B850, and B875 as optical platforms since these CDW's are "locked" in place by the "frozen" phonon mode .
3. Prevent radicals or energy "back flow" from the R/C.

## 3. Coherence, Polariton States and Their Impact on the PLHC. [5, 19, 20]
### a. What is Coherence?

If a single BChl molecule in the ground state absorbs a photon, a ground state negatively charged electron transitions to an excited state called an exciton, and the resulting molecule contains a ground state positively charged "hole" left by the electron. Depending on the material, the excited state might be "tightly" bound (the distance between two charges are about a lattice distance in which case it is called a "Frenkel exciton". If the distance between the hole and electron is much larger than a lattice spacing, the exciton is a "Wannier-Mott exciton". The ground state hole and the excited state electron are both fermions (odd spin), so the combination is now a boson (even spin). If we have a hypothetical box of ideal, spinless bosons close together (higher enough density) so they may interact and the temperature is lowered below $T_c$, they will condense into a BEC (Bose-Einstein Condensation). The density is higher and the temperature is lower to enable a BEC.

A Hamiltonian describing these bosons / excitons is

$$H = \sum_i \frac{\hbar^2 k^2}{2m} \hat{a}_{\vec{k}}^\dagger \hat{a}_{\vec{k}}$$

where $\hat{a}_{\vec{k}}^\dagger, \hat{a}_{\vec{k}}$ are the creation and annihilation operators, and $k$ is the wave vector The distribution number of *bosonic* particles is different from fermions (-1 is changed to +1 for fermions)

$$N_{\vec{k}} = \frac{1}{e^{\beta(\hbar^2 k^2/2m - \mu)} - 1} \qquad (21)$$

with $\mu$, the chemical potential and $\beta = 1/k_B T$, where $k_B$ is the Boltzmann constant. Normally, in the thermodynamic limit (large number of particles), the density of particles is

$$n = \lim_{V \to \infty} \frac{N}{V} = \frac{1}{V} \sum_{\vec{k}} N_{\vec{k}} = \frac{1}{(2\pi)^3} \int d^3k \frac{1}{e^{\beta(\hbar^2 k^2/2m)} - 1} \qquad (22)$$

so the integral becomes

$$n = 2.612 \frac{(mk_B T)^{3/2}}{(2\pi\hbar^2)^{3/2}} \quad \text{or} \quad T_c = \frac{2\pi\hbar^2}{k_B m} \left(\frac{n}{2.612}\right)^{2/3} \qquad (23)$$

However, the integral has not accounted for all the particles since a discrete function (the sum) was replaced by a continuous one (the integral), implying zero states at zero energy. This zero state, $n_0$, must be treated separately so rewriting the density eq. (7)

$$n = \lim_{V \to \infty} \frac{N}{V} = n_0 + \frac{1}{(2\pi)^3} \int d^3k \frac{1}{e^{\beta(\hbar^2 k^2/2m)} - 1} \qquad (22')$$

When a large number of bosons are cooled below $T_c$ (*note the inverse dependence on mass*, eq. (23)), a large fraction of the bosons occupy the lowest state and become coherent which means that the phase, $\varphi$, is no longer random ($0 \to 2\pi$), but becomes a single value and each boson

"phase locks" with all of the other bosons. The phases of the wave functions of all the participating particles are the same. Identifying bosons as excitons would mean that we could diagonalize the exciton density matrix in a new basis and represent all the participating excitons with a single wave function, $\psi(\vec{r},t) = \sqrt{n_0(\vec{r},t)}e^{iS(\vec{r},t)}$, where $n_0$ was defined by eq. (22') and $S(r,t)$ is the common phase. In a "normal" state the phases of a wave function are an arbitrary value $\varphi$ value that varies from $0 \to 2\pi$.. In a BEC this phase value is arbitrary, but a single value; the inherent electromagnetic field symmetry has been "broken". Further, the chemical potential has now become zero, so the "resistance" to adding or removing a particle is now zero (see fig. 8).

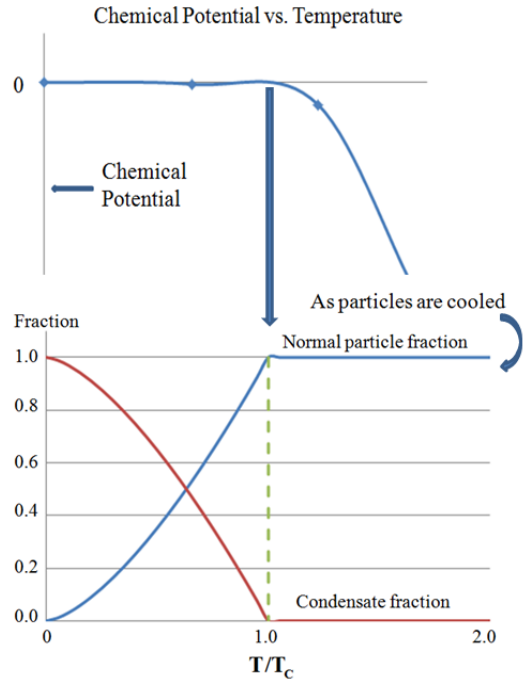

Figure 8. When the chemical potential reaches zero, $T_c$ is defined and BEC begins. Thus, the energy cost to add or remove a particle in an ideal BEC is zero. Note: BEC's (including BCS) are non-perturbative solutions; i.e. there is a singularity in the solution.

Notice in eq. (23) the temperature and mass are inversely related. The physical significance of this relationship is that the particles can be represented by the thermal de Broglie wavelength

$$\lambda_{dB} = \left(\frac{2\pi\hbar^2}{mk_B T}\right)^{1/2}$$

that measures the extent of the quantum mechanical wave packets (invert LHS eq.23). As the temperature is lowered, the wave packets begin to overlap and probability of finding multiple bosons in the ground state increases dramatically, a phase transition results and the bosons form a BEC. When the mean de Broglie thermal wave length of the particles approaches unity, quantum effects characterize the system. Conversely, when $n\lambda^3 \to 0$, the system goes smoothly to the classical values. A BEC is the result of quantum statistics as opposed to inter-particle interactions. Nozieres argues that in bosons exchange interactions even work to prevent fragmentation of the condensate [24]. The BEC phase transition results from the particle wavelength spanning the interparticle distance, roughly the same point when exchange effects become important. If we assume that our excitons are identical bosons with the same wave function and if they are and at the "right" temperature (usually low), these entities can interact if the density is high enough. The results in some materials are coherent exciton waves [25] in an inorganic semiconductor.

If energy transfer is of the Förster type transfer such as a "dimer resonance", etc., movement of energy can dissipate quickly since the vibrational components are directly connected to the continuum; return is highly improbable even with the paucity of continuum states. An exciton has a mass roughly $0.2 m_e$, in general too low to reach room temperature BEC condensation. The "extended Förster resonance" still has a connection to a vibrational continuum, and thus similar restrictions. A bosonic exciton can combine with a photon to create a polariton which does have a small enough mass ($10^{-4} m_e$) and is capable of room temperature condensation. A polariton can also capture multiple photons.

**b. Polariton States [23, 24, 25, 19]** A polariton is a two-level state created by the superposition of a photon and another boson with a dipole such as a phonon, exciton or plasmon. The interaction Hamiltonian is the dot product of the dipole energy (exciton) in the polarization electric field of a photon, $\vec{E}$,

$$H_{int} = -\int d^3 r \vec{P} \cdot \vec{E}$$

Polaritons have been known for years, but the difficulty in studying them disappeared with the creation of microcavities to trap photons for a much longer time. This technique enabled a much larger density, further increased by the recognition that the large oscillator strength of organic materials leads to a much larger Rabi separation between the levels (1.2 eV). Both of these advances allow a macroscopic number of polaritons to occupy the same state so a Bose-Einstein condensate (BEC) is formed despite the fact that these quasiparticles[*] are *inherently in non-equilibrium states*. The PLHC can absorption many additional photons to create more excitons, but two excitons in a limited space will interact annihilate each other, thus coherence is not favored. An alternative solution is the creation of a coherent exciton-polariton, a combination of two bosons, an exciton and a photon where the combined mass is much lower (Fig 9).

**b.1. Normal Emissions.** Einstein showed that if matter and radiation are to be in thermal equilibrium, the process of spontaneous emission has to occur. Excited atoms must radiate with a collective lifetime $\tau$. How does one atom in a collection of atoms "know" when it is supposed to radiate? Spontaneous emission is not a process that is exclusive property of the collection of atoms, but one of the atom-vacuum system. The irreversibility of the emission is due to an infinity of available vacuum states. The probability of photon emission per unit time $P_e(t) = e^{-\Gamma_0 t}$ where probability $\Gamma_0$ can be calculated by Fermi's "golden rule". The vacuum field is part of the system is a huge collection of oscillators in which the excitation rapidly and irreversibly decays.

---

[*] A quasiparticle can be described as a real particle and its environment whose behavior is influenced it; also called a "dressed particle."

**b. Quantum Cavities Emissions [26, 27, 28].** In the late 1946's Purcell made an observation that placing an excited atom or molecule in a cavity could either enhance or diminish spontaneous emission [26]. In the late 1980's technology was advanced enough to create a useful cavity and cavity quantum electrodynamics was born in a series of experiments [27, 28]. In a cavity the mode structure of the vacuum is significantly reduced if the cavity size, d, is comparable to the wave length. Further, if an excited atom or molecule has *an electric dipole oscillating parallel to the cavity "wall"* (mirrors or Distributed Bragg Reflectors (DBR)), its lifetime can become infinitely long if $\lambda < d/2$. The B850 complex structure (Fig 10A (top)) meets this criteria; is it just a nano-cavity or nano-resonator in conjunction with other B850's surrounding B875? Lifetimes have been extended from 13 to 20 times the free space values in a Rydberg state with a small waveguide.

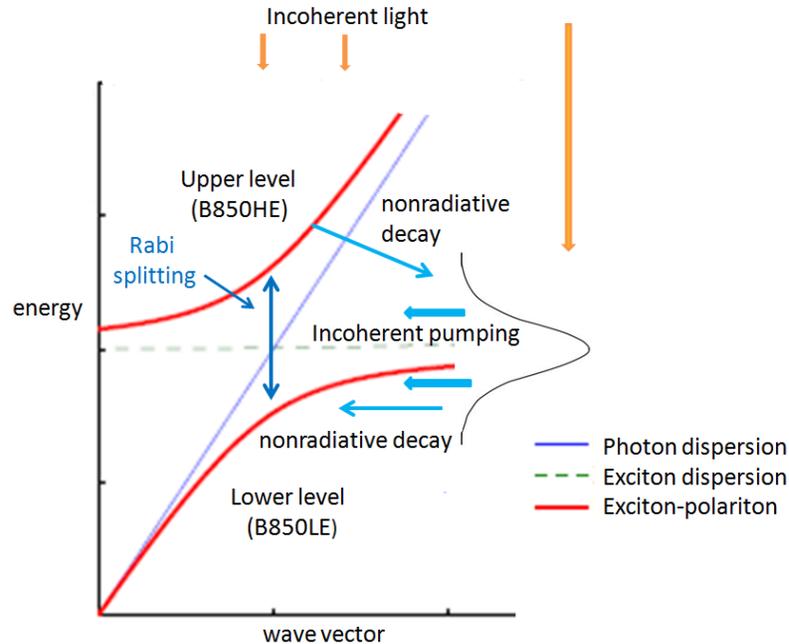

Figure 9. An exciton comprised of an electron and a hole is pulled in opposite directions by an electric field affecting the dielectric constant of the medium. The resonance frequency of the exciton is not constant, resulting in variable branches of the exciton-polariton, the strength of which is measured by the Rabi splitting between the two branches. The stronger the oscillator strength of the exciton, the larger the splitting. The eigen-modes mix only in the region of the crossover.

### c. Polariton Influence on PLHC; Coupling Between an Exciton and Photons in a B850 Cavity.
The coupling between excitons and photons is strong in the region of the crossover so

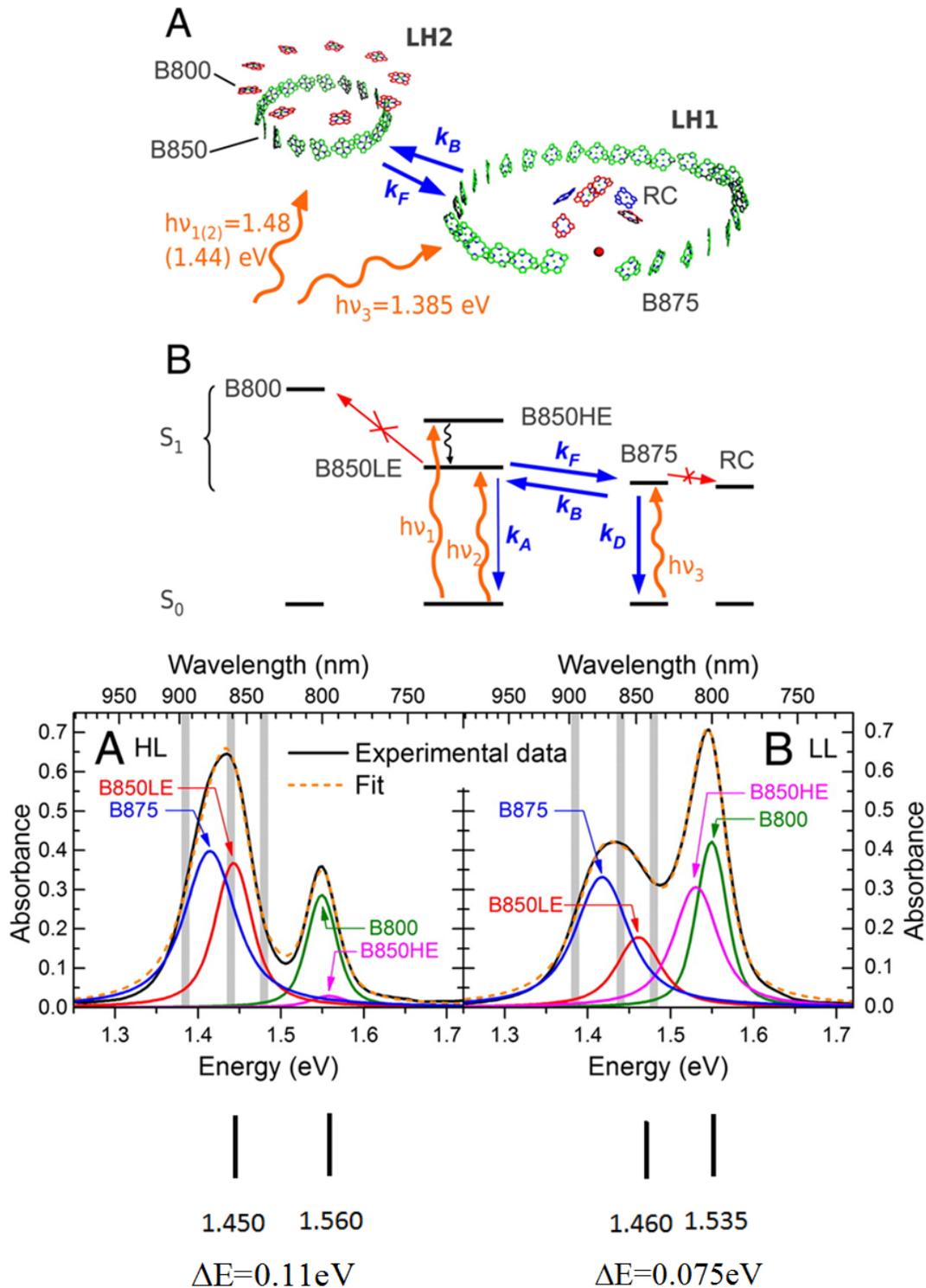

Figure 10. Schematic of the BChls in LH2 and LH1/RC complexes from R. palustris.

Fig 10C. Energy splits between B850LE and B850HE at HL and LL

Figures 4A and 4B used with permission of the National Academy of Science

the exciton-polaritons are the dominant eigenstates. These eigenstates do not decay into other states unless there is an additional interaction as their lifetimes are determined by their path length in the medium. The eigenstates of the polaritons are doublets separated by the value of the Rabi splitting (fig 9). The coupling between excitons and photons can be categorized in terms of coupling strength, weak and strong. Weak coupling can be treated perturbatively as the emission spatial and spectral distribution may be altered, but the exciton dynamics is essentially intact. In inorganic quantum wells this splitting is about 3 – 10 meV, so experiments must be done at very low temperatures. All of the original studies focused on inorganic semiconductors which have relatively small Rabi splitting $(\hbar\Omega \sim 5 meV)$ at 5K indicating the weak regime.. The first observation of a cavity exciton-polariton was by Weisbuch et al [29]. The cavity contained five quantum wells containing an inorganic semiconductor with a Rabi splitting of meV's. The reference contain figures that are instructive concerning the cavity, DBR's, and tuning/detuning the photon energy to find resonance.

**Application of Coherent Theory / Polaritons to Experimental Results. a. General Discussion.** The near perfect energy transfer in the PHLC from absorption, energy transfer over significant distances and finally to charge separation is legendary. Our initial thoughts were focused on something more than exciton hopping and/or tunneling to the extended Förster model. However, several examples such as quantum "beats" [30, 31] had made it obvious that many portions of the PHLC were coherent, but what are the interacting states? The often-mentioned impact of "the dimer" was discussed, but later ignored and fermion symmetry operations were performed by us and others as if the dimers did not exist. Most analyses were made with the fermion states being the dominant ones so a "stack" of fermionic states with +/- k components resulting. For each +k component, there is a negative one, so adding the two together results in all states being zero angular momentum, i.e. $\Delta p = 0$; these states are bosons. This would predict that cyclical molecules like benzene and coronene (same number of pi electrons as a BChl) could contain paired electrons (pseudobosons) in their ground state [32]. Recent experimental results confirm that both benzene and coronene photoemit a charged pseudo boson [33]. Alternative methods of diagonalization are the Lancos technique or Coulson's axial symmetry method.

The individual Mg-BChl molecules also should contain Peierls distortions [11], as well as $\beta-$carotene [34, 35, 36], the latter due to the alternating carbon single and double bonds. We can also make a case for the excited state(s) to be coherent. In principle these paired $(\Delta p = 0)$ exciton states on a ring can be diagonalized into a single coherent "macrostate", $\Psi = \sqrt{\rho}e^{i\varphi}$ spanning the entire B850 and B875 rings. While multiple excitons would quickly annihilate each other, coherent, RT polaritons will preferentially be generated.

Upon receiving Fig 1, we recognized the Peierls' distortion indicating the ground state of B850 and B875 is fully coherent at room temperature. However, there seems to be considerable coherence in the entire PLHC so we will suggest possible coherences. We begin with the backflow energy diagram from [1] (Fig 10).

**c. Electronic Energy Transfer.** In addition to the near perfect forward energy transfer, another component of the energy transfer between LH2 and LH1 is the ability of *uphill* energy transfer from B875 to B850 [1, 37, 38]. In various species of bacteria, the organism modifies the spectral intensity and disorder based on the amount of light present when the samples are grown which influences the type and energy of the LH2 complexes synthesized. Lüer et al have devised an experimental scheme to compare high and low light conditions (HL and LL). Examination of Fig 10A and 10B, reproduced from [1] illustrates the significant intensity splits of B850 into high energy (B850HE) and low energy (B850LE) bands to describe the spectra properly [36, 37]. Presumably, the split is attributed to disorder caused by a heterogeneous, non-symmetric composition of apoproteins on a timescale with a picosecond exciton relaxation. Normally exciton-exciton splitting is nearly symmetrical with an energy barycenter reflective of the initial energy of the two excitons [12].

The B850HE and B850LE energy difference is non-linear, indicating the possible presence of a polariton composition of these energy levels. We assume that the initial B850 exciton has absorbed additional photons thereby forming a polariton, Fig 11. Since a polariton has the ability to extend its length to the order of the size of the microcavity mode, we suggest the B850 polariton can easily span the B850 ring circumference, can further use the "sunflower" expansion mode of circular polaritons; the B850 ring polariton then likely interacts with the B800 molecules [38]. This mechanism could explain the peculiar relationship that B800 seems to have relative to B850 BChl's such that when two units of B850 are added, presumably to maintain the Peierls' distortion, a single B800 resides between two B850 units to enhance light capture. These B800 units provide an incoherent or a coherent reservoir of energy to the B850UL [25].

**e. "Uphill and Elsewhere" Energy Transfers. (Figure 11) 1.** The polariton's levels (B850 LE and HE) resonantly exchange energy. In HL and in addition to the normal light absorption, the EET proposed above might follow also follow a route where B800 absorbs light and transfers energy to the extended B850HE branch of the polariton that exchanges energy with the B850LE and ultimately transfers it to B875. In HL to prevent the possibility of B875 being damaged, it shifts energy to B850LE which has shifted close to the B875 energy. This could be possible if the B875 wave oscillations of the highly populated excited state can provide a transient higher energy for B875 and the B850LE a transient lower wave in a similar manner, equal or lower energy than transient B875; uphill EET can take place. This process might also take place between other B850 complexes surrounding the B875 chain. Thus, backward flow might be a mechanism for energy storage. Further, the B875 could use the gap wave-like nature (oscillation) to transfer energy to the RC. Lastly, the possibility of a B850 coherent polariton extended its length half-way across the LH1-B875 complex would suggest that the eight B850 complexes surrounding the LH1-RC could provide an extensive energy supply as well as ample storage capability using a modified Jaynes-Cummins effect [39, 40, 41] which can become spectroscopically dark (see Figure 12).

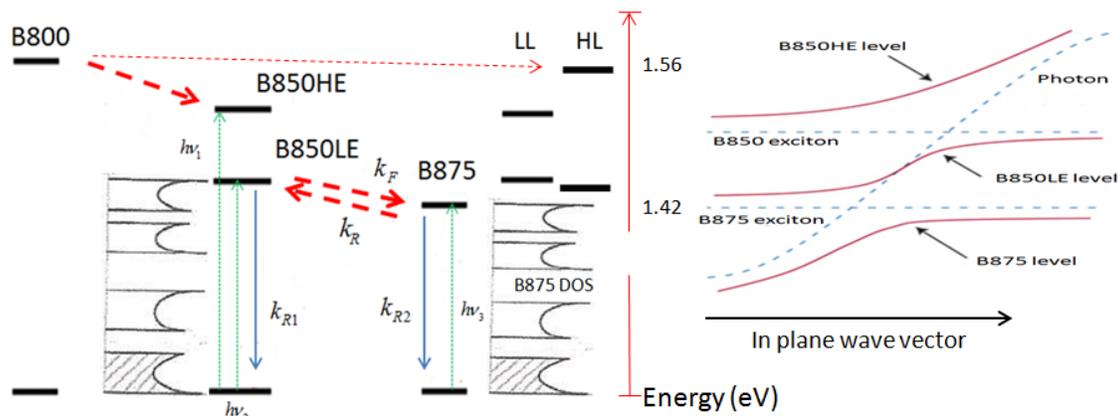

Figure 11. Excited state energy levels and band gaps for B800 molecules (left), then eight B850's surrounding the B875/RC along with possible polariton structures for these states. Most bosons are restricted to the region of $\Delta p \sim 0$; polaritons can have considerable freedom in this regard, covering distances to $50\mu m$ or more [42].

___________________________________________________________________________

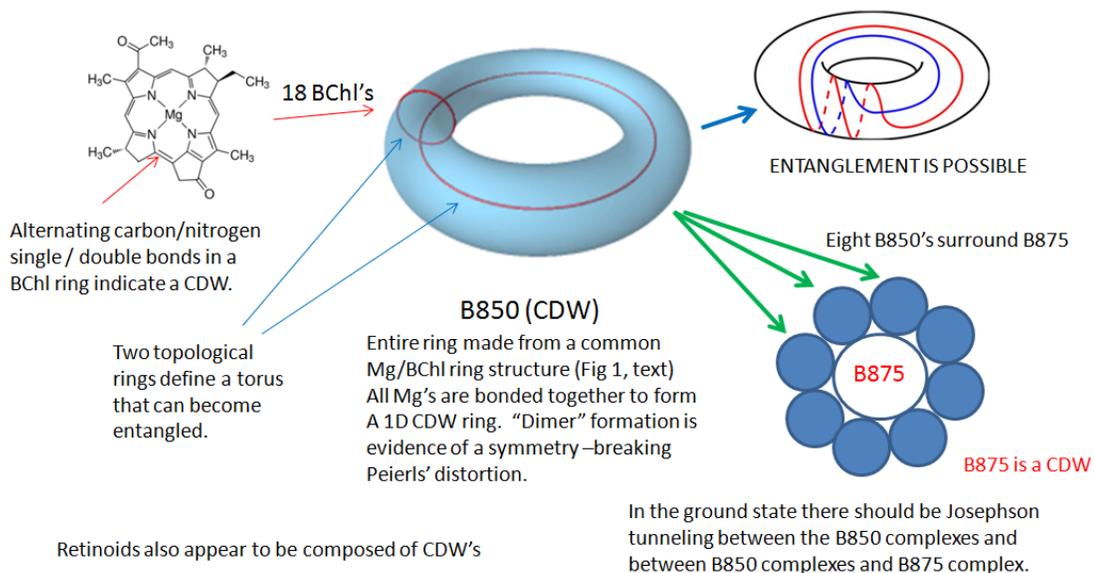

Figure 12. Circular CDW's in a magnetic field have been shown to quantize the magnetic flux in units of $\Phi = nh/2e = n\Phi_0$ where n is the number of flux. The Meissner effect is a rotation of charge in a SC or CDW that opposes the external flux, permitting only the flux in the fundamental units of $\Phi_0$ (or $h/2e$). In principle BChl could generate an n, a B850 complex, n', and the eight B850 surrounding B875, another n'' (right, top and b).

**4. Possible Energy Storage Mechanism - Jaynes-Cummings Model.** The Jaynes-Cummings model (JCM) is a model of a two level system interacting with a quantized cavity mode. The interesting point is that the B850LE and HE appear to be a polariton and if electronic energy backflow is a storage mechanism, the JCM would be the likely model. Essentially, since the quantum number of an imdivual wave is a single number, but the average number of particles is variable since the phase is fixed, there exists interference between the actual and average that generates beats. The result is that the population of a two level system (polariton) can become spectroscopically dark. The JCM for polaritons, described in figure 12, has been observed in polaritons [43, 44].

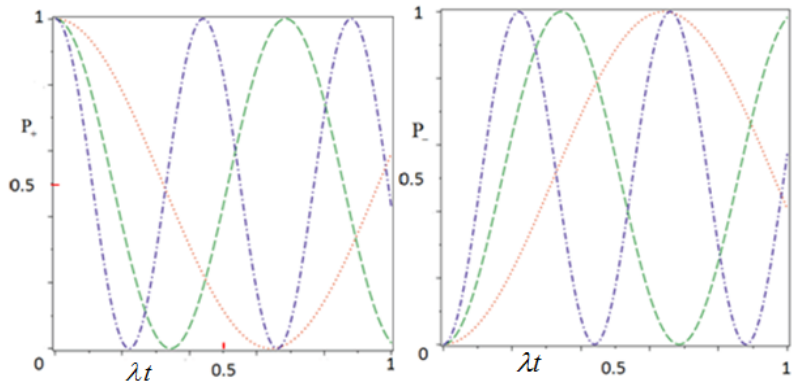

A. Probability Amplitude Oscillations – n=5 (red); n=10 (green); n=50 (blue)

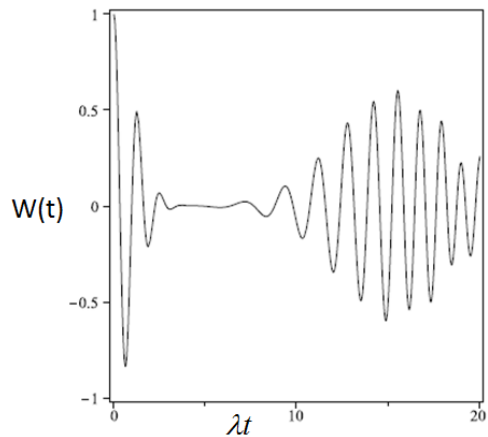

B.

$$W(t) = P_+(t) - P_-(t) = e^{-N}\sum_{n=0}^{\infty}\frac{N^n}{n!}\cos\left(2\lambda\sqrt{n+1}t\right)$$

Figure 13. Coherence implies a single phase, so the uncertainty principle requires a distribution in the number of particles. If n is the exact number of photons and N is the average number, a perfect two-level system will show "beats", "collapse", and "revival" if there is no decoherence. This process is a possible photosynthesis storage system.

**5. Josephson Junctions (JJ).** Shortly after the BCS theory was published, Josephson predicted that if two superconductors were separated by an insulating barrier, the SC current (Cooper pairs) can tunnel through the normal barrier if it is not too thick [16]. The effect is based on the phase difference between the two superconductors which creates a potential difference,

$$\partial(\Delta S)/\partial t = 2eV/\hbar ,\qquad(8)$$

the *current is powered by the phase difference.* Writing the condensate wave function as $\psi(\vec{r},t) = \sqrt{n_0(\vec{r},t)}e^{iS(\vec{r},t)}$ to emphasize the collective mode and assuming the phase and amplitude are allowed to vary over long distances, it can be shown that the current is proportional to the gradient of the phase, $\vec{j} = \frac{\hbar}{m} n_0 \nabla S$; a difference in phase will generate a superfluid flow. The condensate wave function can also be used to calculate the energy associated with London's "phase stiffness" $(\rho_s)$ can be expressed as $H_s = \rho_s \int d^3r (\nabla S)^2$ which measures resistance to twisting. Josephson wrote a very clear review [16] shortly after his extraordinary discovery which is currently a major field of research.

There are two possible applications of JJ in addition to CDW's. Oscillations between two different frequency condensates, presumably separated in two different wells with a gap in between, will create "beats". These beats are analogous to the ac Josephson oscillation effect due to the difference in chemical potential. If one well has a low population and therefore, a large $\Delta\theta$, the Josephson current rapidly replenishes it. This type of response could be operative in the B850LE/B875 "uphill" response as well as the B875 transfer of energy to the RC [38].

**6. Conclusions and Speculations.**

1. The symmetry breaking generating the BChl "dimer" in the crystal structure of CDW's in B850 and B875 complexes conclusively establishes that the ground state is coherent. The coherent phonon mode that spans the B850 ring results a highly stable optical platform.

2. The energy gaps created by the symmetry of the band structure of the B850 rings and the CDW gap drastically reduce the number of vibrational modes for energy relaxation from the excited states.

3. The polariton model offers very fast (picosecond) energy transfer kinetics and a mechanism for coherent capture of incoherent absorbed light. (possibly B800 to B850).

4. All of the surrounding B850 rings and the B875 complex "communicate" with each other in the ground state by Josephson junction tunneling to creating a foundation for a PLHC "supercomplex". We propose that the most important energy transfer excited states of the B850 and B875 complexes are coherent polaritons, an excited state created by mixing an exciton and photons. We propose that this total structure is coherent both in the ground and excited states.

5. The polariton model offers very fast (picosecond) energy transfer kinetics and a mechanism for coherent capture of incoherent absorbed light. (possibly B800 to B850).

6. The Jaynes-Cummings Model applied to a polariton offers a mechanism for storage of

photons in a two level system; this state can become "dark" by oscillating from "bright" to "dark" and back.

7. We speculate that the arrangement of the LH2 and LH1 could have properties of a nanocavity. the eight B850 and B875 might act as a toroidal structure and a resonator surrounding the RC, respectively.

8. Since the B850 complex ground state is a coherent phonon state around the ring, application of a magnetic field ought to produce a Meissner effect that might be measurable. The Meissner effect only allows integer amounts of magnetic flux through the ring, i.e. the flux is quantized.